# Ultracompact plasmonic racetrack resonators in metal-insulator-metal waveguides

Z. Han

*Department of electrical and computer engineering, University of Alberta, Edmonton,*

*T6G2V4, Canada*

**Abstract**-Among various plasmonic waveguides, the metal-insulator-metal (MIM) type is the most promising for true subwavelength photonic integration. To date, many photonic devices based on MIM waveguides have been investigated, including resonators. However, most of the reported MIM ring resonators suffer from low extinction ratios and the reasons are unexplored in the literature. In this paper, we present a comprehensive analysis of the intrinsic reasons for the low performance of MIM ring resonators, and give the analytical transmission relation for a universal all-pass ring resonator with coupling loss. Based on the analysis we propose plasmonic racetrack resonators in MIM waveguides and show that the performance can be greatly improved.

# 1. Introduction

Plasmonic waveguides have recently attracted much attention due to their ability to confine light beyond the diffraction limit [1], therefore potentially enabling device integration on a scale which is not accessible with conventional dielectric waveguides or photonic crystal waveguides. To date, various plasmonic waveguide structures have been proposed and investigated for planar photonic integration [2-5]. In general most plasmonic waveguides can be classified into three basic types: insulator-metal-insulator (IMI), insulator-metal (IM) and metal-insulator-metal (MIM). In the optical regime, especially in the infrared, light in IMI and IM types of plasmonic waveguides penetrates into the insulator over several wavelengths, so the mode profile in these two types is still quite large. In contrast, the MIM type can provide true sub diffraction confinement when the width of the insulator is very small, and in principle, is not limited by a cut off. Thus the MIM type of plasmonic waveguides is the ideal candidate for true subwavelength photonic integration, albeit the loss is larger than the other two types.

To realize subwavelength photonic integration based on the plasmonic platform, guidance of light with nanoscale confinement is required, but more importantly some functionalities, e.g. filtering, routing and multiplexing, have to be realized on the same platform. One of the devices that can provide such functionalities is a resonator, which can filter specific wavelengths from an input signal. Filters based on Bragg gratings in MIM waveguides have been investigated in the literature [6]; however, these devices require a relatively large length because a large number of Bragg periods is needed. The recently proposed tooth-shaped plasmonic waveguide filter is shorter [7], but the bandwidth is too large for filtering applications.

Considering that light can propagate through sharp bends in MIM waveguides with high transmission [3] [8], MIM waveguides are ideal candidates for constructing ring resonators, in which the bending loss is negligible and ring resonators with subwavelength radii can be achieved. Using MIM waveguides, resonators with very short round trip length $L_{rt}$ can be realized. A shorter $L_{rt}$ is preferable for plasmonic ring resonators due to a lower propagation loss inside the resonator. In addition, a short $L_{rt}$ results in a large free spectral range (FSR), which is desirable for filters in wavelength-division-multiplexing applications [9]. In ref [10] plasmonic ring resonators based on the regular circular geometry were theoretically investigated and ring resonators based on channel plasmonic waveguides were reported in ref [5]. However, all these resonators suffer from a relatively low extinction ratio. To show one of these results, Fig.1 gives the calculated transmission spectrum for a MIM ring resonator with a radius of 500nm evanescently coupled to a MIM bus waveguide. The spectrum was obtained using the finite-difference time-domain (FDTD) method [11] with convolutional perfectly matched layer (CPML) boundary conditions. The index of the insulator is assumed to be 1.45, which is the typical index of $SiO_2$ or some polymers in the infrared region. The metal is silver, with the permittivity described by Drude model $\varepsilon_r = \varepsilon_\infty - \omega_p^2/(\omega^2 + j\gamma\omega)$, with $\varepsilon_\infty$ = 3.7, $\omega_p$ = 9.1eV and $\gamma$ = 0.018eV, which were obtained by fitting the experimental data [12] for silver permittivity in the infrared frequencies. Both the bus waveguide and the ring resonator have an insulator width of 100nm and the gap between them is 25nm, which should be realizable and controllable with state-of-the-art electron beam lithography or focused ion beam lithography.

From Fig.1 we can see that for resonant wavelengths in the range between 1μm and 2.25 μm, the extinction ratio for this circular MIM ring resonator is very small, less than 5dB. Such a low extinction ratio limits the application of this ring resonator; however, the reasons for the low extinction ratios have not been analyzed in the literature. In section 2 of this paper, we will give a comprehensive analysis to show why the extinction ratio is so small and how it can be improved. We note that although the transmission of an all pass ring resonator has been given in ref [13], it was assumed therein that there is no loss in the coupling region, which is not true in the plasmonic device. In section 2 we present an analysis for a universal all pass ring resonator where loss exists in the coupling region. Based on the analysis, we propose in section 3 a MIM racetrack resonator with significantly improved extinction performance. We note that although the analysis and design presented in section 3 is for MIM waveguides, it can also be extended to other plasmonic waveguide systems. Moreover, the analysis in section 2 is not restricted to MIM resonators, but is applicable to any resonators and can be used as a guideline for designing of all types of plasmonic resonators with high extinction ratios. The paper concludes in section 4.

## 2. Analysis of a universal all-pass ring resonator

The schematic of a universal all-pass ring resonator is shown in Fig.2. When the coupling region has loss and non-zero physical length, the coupling coefficient $\kappa$ and transmission coefficient $\tau = |\tau|e^{i\phi_\tau}$ are both complex numbers, and we assume $|\tau|^2 + |\kappa|^2 = \beta$, where $\beta$ may not be equal to 1. The coupling relation can be described by the matrix:

$$\begin{vmatrix} b_1 \\ b_2 \end{vmatrix} = \begin{vmatrix} \tau & \kappa \\ -\kappa^* & \tau^* \end{vmatrix} \begin{vmatrix} a_1 \\ a_2 \end{vmatrix} \tag{1}$$

and the transmission around the resonator is given by

$$a_2 = \alpha e^{i\theta} b_2 \tag{2}$$

where the real number $\alpha$ is the inner circulation factor describing the internal loss, with α = 1 when there is no internal loss, and $\theta$ is the round trip phase.

From (1) and (2), we can get:

$$b_1 = \frac{-\alpha\beta + \tau e^{-i\theta}}{-\alpha\tau^* + e^{-i\theta}} a_1 \qquad a_2 = \frac{-\alpha\kappa^*}{-\alpha\tau^* + e^{-i\theta}} a_1$$

So the transmission is:

$$T = \frac{|b_1|^2}{|a_1|^2} = \frac{\beta^2\alpha^2 + |\tau|^2 - 2\alpha\beta|\tau|\cos(\theta + \phi_\tau)}{1 + \alpha^2|\tau|^2 - 2\alpha|\tau|\cos(\theta + \phi_\tau)} \tag{3}$$

From equation (3) one obtains the transmission at resonance ($\theta + \phi_\tau = 2m\pi$, m is an integer) as:

$$T = \frac{(\alpha\beta - |\tau|)^2}{(1 - \alpha|\tau|)^2} \tag{4}$$

So the condition for critical coupling in ref [13] is now changed to $\alpha\beta = |\tau|$, which determines the extinction ratio for a universal all-pass resonator. Compared to dielectric ring resonators, plasmonic waveguides suffer from more loss in the ring and there is also some loss in the coupling region, so that α and β are relatively smaller than in the dielectric structures (β=1.0 for dielectric waveguide). Furthermore, $|\tau|$ is larger in the plasmonic device because more power goes through the straight bus waveguide due to the weak coupling. These two conditions make the deviation between αβ and $|\tau|$ for a plasmonic resonator larger than that for dielectric resonators. This is the origin of the low extinction ratio in MIM plasmonic resonators. In the following part, we will give specific values of αβ and $|\tau|$ to show the deviation between them. One can also conclude that for plasmonic resonators, a stronger coupling is required to achieve a high extinction ratio due to the loss.

The key to increasing the extinction ratio of MIM ring resonators is to reduce the deviation between αβ and $|\tau|$ at resonant wavelengths. Some approaches, e.g. aperture-coupling [14], have been proposed to meet this requirement. Since light coupling between two waveguides is proportional to the coupling length, we propose in this paper MIM ring resonators with increased coupling length, i.e. plasmonic racetrack resonators based on MIM waveguides.

We first show how the deviation between αβ and $|\tau|$ can be reduced in the racetrack resonator at 1550nm. For simplicity, we assume the effective index of the bent MIM waveguide is the same as that of the straight MIM waveguide, so α equals to $e^{-\frac{2\pi}{\lambda_0}n"L_{rt}}$, where $\lambda_0$ = 1550nm is the free space wavelength, $n"$ is the imaginary part of the effective index in the MIM waveguides, $L_{rt} = 2(\pi R + L)$ is the round trip length of the racetrack resonator where R is the radius of the ring and $L$ is the length of the coupling region. By solving the dispersion equation of a MIM waveguide at 1550nm with the insulator width set at 100nm and its index equal to 1.45, we obtain $n_{eff} = n' + n"j = 1.747 + 0.00327j$, which corresponds to the propagation length in both the resonator and bus waveguide to be 37.72μm (the loss is 0.115dB/μm). The relation between $\alpha$ in the racetrack resonator and coupling region length $L$ can be obtained. To calculate τ, κ and β, we simulate the coupling region separately as shown in the inset in Fig. 3(a) using FDTD. The material parameters are the same as given above. A continuous wave at 1550nm is launched from the left port of the straight waveguide and the simulation is run until the fields become stable. The magnetic fields at planes A and B are used to calculate τ, κ with the magnetic field at plane A without the upper waveguide used as the reference. Fig. 3(a) gives the result for τ and κ as well as αβ for different coupling lengths.

From the result we can see that as the length increases, the coupling between the two waveguide becomes stronger and the deviation between τ and αβ decreases. When the length L reaches approximately 670nm, τ equals to αβ and when L exceeds 670nm, the deviation between τ and αβ starts to increase again, resulting in a state of "over coupling". Fig. 3(a) also shows that when L equals 0, which corresponds to the case for the coupling in a circular plasmonic resonator as shown in the inset of Fig. 1, τ is 0.9922 while αβ is 0.9551. Thus according to equation (4), T equals to 59.1% or -2.3 dB at resonance. This result is consistent with the result shown in Fig. 1 obtained with the FDTD method.

**3. Proposal and numerical results.**

The result in Fig.3 (a) shows that when the length of the coupling region approaches 670nm, a state close to critical coupling can be realized and this can be used to design a resonator with a much higher extinction ratio. The schematic of such a plasmonic resonator with increased coupling length, or racetrack plasmonic resonator, is shown as inset in Fig 3(b). The FDTD method is again used to simulate the structure and the calculated result is shown in Fig. 3(b). One sees that the extinction ratio around 1.55μm has been significantly increased from less than 3dB to around 35dB. It is also seen in Fig. 3(b) that the extinction ratio at the other resonant wavelengths within the range between 1μm and 2.25μm is also increased, although not as significant as it is for the resonant wavelength around 1.55 μm, indicating the condition for critical coupling is less fulfilled for those wavelengths.

Considering that the coupling is weaker for longer wavelengths, we believe that it is under-coupling ($\alpha\beta<\tau$) for wavelengths longer than 1.6 μm and over-coupling ($\alpha\beta>\tau$) for wavelengths shorter than 1.4 μm.

Compared to the value $\Delta\lambda_{FWHM}$ (full width at half maximum) of 8.1nm obtained for the device in Fig. 1 at the resonant wavelength 1.386μm, the value $\Delta\lambda_{FWHM}$ obtained for the racetrack resonator is about 13.6nm around the resonant wavelength 1.565μm. Considering that $\Delta\lambda_{FWHM} \approx \frac{\lambda^2}{\pi n_{eff} L_{rt}} \frac{1-\alpha\tau}{\sqrt{\alpha\tau}}$ [15], the slight increase in $\Delta\lambda_{FWHM}$ of the racetrack resonator indicates that the decrease in $\tau$ plays a more important role than the increase in $L_{rt}$. We also find that the FSR at around 1.55μm in the racetrack resonator (387.1nm) is smaller than that in the circular resonator (453.6nm), which we believe is mainly due to an increase in $L_{rt}$. The loaded quality factor for the racetrack resonator at 1.565μm is found to be around 115. The unloaded or intrinsic quality factor for the MIM racetrack resonator at 1.55μm is calculated to be 267 from the following equation [16]:

$$Q_{int} = \frac{n'}{2n''} \qquad (5)$$

The relative low quality factor is mainly due to the intrinsic loss in the plasmonic waveguide, which is a common problem for all plasmonic resonator systems and we hope this problem can be solved by introducing e.g. gain medium into the plasmonic waveguides. Compared to dielectric ring resonators, e.g. the smallest silicon-on-insulator ring resonator reported to date in the literature with a radius of 1.5μm [17] which has a FSR of 52nm, the proposed MIM racetrack resonator has a smaller quality factor, but has a much smaller bending radius and larger FSR while retaining a comparable extinction ratio. We note that the bending radius can be even further reduced due to the high transmission of light through sharp bends in MIM waveguides.

One may also notice that the extinction ratio for the resonant wavelength around 1.55 μm is only 34dB while from equation (4) this number should be infinity. Besides the numerical errors, the main reason for this is because the results for coupling shown in Fig. 3(a) is calculated for the 1.55μm wavelength while the resonant wavelength in Fig. 3(b) is precisely 1.565μm. For 1.565μm $\alpha\beta$ and $\tau$ are not exactly equal. The small difference slightly degrades the extinction ratio for 1.565μm; however this can be improved by carefully designing the plasmonic racetrack resonator geometry to fulfill both the phase condition and critical coupling condition at the same time. For the result shown in Fig. 3(b), we note that the length 670nm is still relatively short compared to the resonant wavelength and the footprint of the plasmonic racetrack resonator is still very small.

Fig.4 (a) gives a snapshot of the magnetic field profile in the plasmonic racetrack resonator at the resonant wavelength 1.565μm. As can be seen in the figure, the field in the output port of the bus waveguide is very weak, showing that complete destructive interference is obtained at the output port. To assess the sensitivity of the extinction ratio to variations in the length of the coupling region, we performed a set of simulations in which the length of the coupling region has some deviations from 670nm. Fig. 4(b) shows the extinction ratios for different coupling length deviations at around 1.55μm wavelength. Note that the resonant wavelengths have changed a bit from 1.565μm because $L_{rt}$ has changed. As is seen, the extinction ratio is slightly degraded, which can be easily understood because the coupling is deviated from the critical coupling discussed in section 2. However, the extinction ratio is still much higher (>20dB) than the result obtained for the circular MIM resonators.

**4. Conclusion**

In summary, we presented in this paper a comprehensive analysis to elucidate the cause for the poor performance of circular plasmonic ring resonators, gave the transmission relation in a universal all-pass resonator where coupling loss exists and then proposed and investigated the racetrack MIM resonators. The simulation results show that the proposed racetrack resonators have significant improvement in extinction ratio over the circular MIM ring resonator. Although we focus on plasmonic resonators based on MIM waveguides as an example, our analysis can be used as a guideline for the design of other kinds of high performance plasmonic resonators. Besides the all-pass resonator, the idea of racetrack MIM resonators can be extended to other resonator systems, e.g. add-drop plasmonic resonator and find extensive applications in the field of plasmonic devices. These devices will help realize more functionality on the plasmonic platform.

**Acknowledgments**

This work was supported by the Natural Sciences and Engineering Research Council of Canada. Z. Han also thanks Marcelo Wu, David Perron and Dr. Vien Van for helps.

**List of Figures:**

Fig.1 Calculated transmission spectrum with FDTD for a plasmonic ring resonator; Inset: schematic of a MIM ring resonator side coupled to a MIM bus waveguide, yellow region: metal and brown region: insulator. R: 500nm, G: 25nm and W: 100nm.

Fig.2 Schematic of a universal all-pass ring resonator with coupling loss.

Fig. 3 (a) the magnitudes of transmission coefficient and coupling coefficient and versus the length of the coupling region; Inset: Schematic of the coupling region, yellow region: metal and brown region: insulator. R: 500nm, G: 25nm and W: 100nm; (b) Calculated transmission spectrum with FDTD method for the plasmonic racetrack resonator; Inset: Schematic of the plasmonic racetrack resonator, brown region: metal, brown region: insulator. R: 500nm, L: 670nm, W: 100nm and G: 25nm.

Fig. 4 (a) the magnetic field profile for the wavelength 1565nm; (b) Extinction ratios versus the length deviations of coupling region.

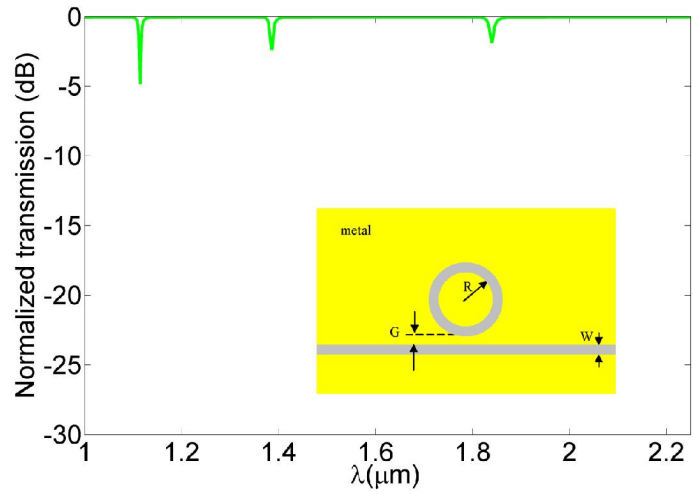

Fig.1 Calculated transmission spectrum with FDTD for a plasmonic ring resonator; Inset: schematic of a MIM ring resonator side coupled to a MIM bus waveguide, yellow region: metal and brown region: insulator. R: 500nm, G: 25nm and W: 100nm.

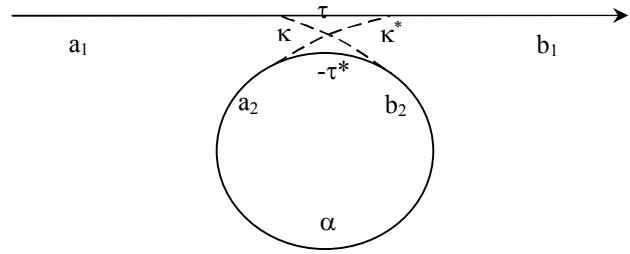

Fig.2 Schematic of a universal all-pass ring resonator with coupling loss.

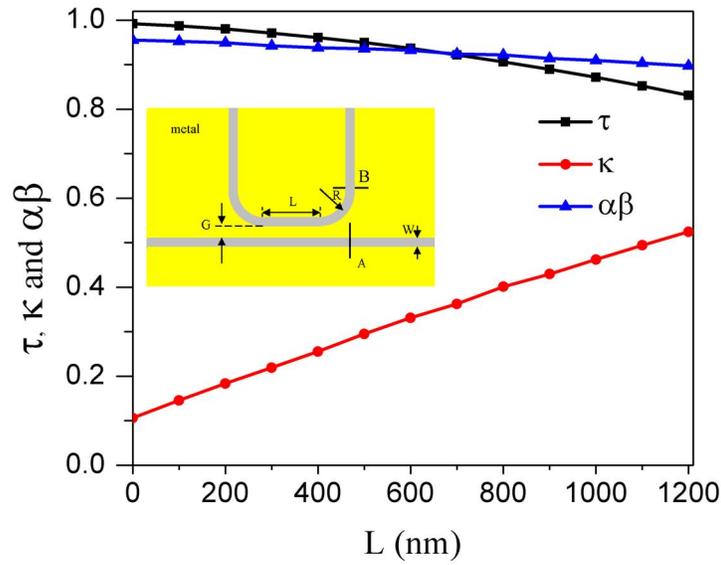

(a)

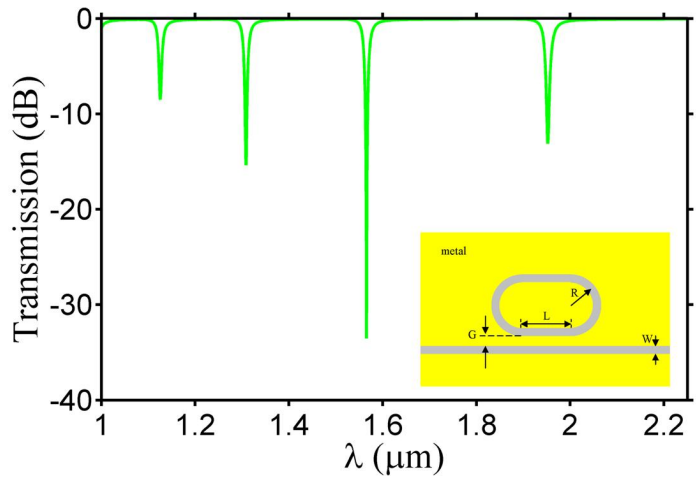

(b)

Fig. 3 (a) the magnitudes of transmission coefficient $\tau$ and coupling coefficient $\kappa$ and $\alpha\beta$ versus the length of the coupling region; Inset: Schematic of the coupling region, yellow region: metal and brown region: insulator. R: 500nm, G: 25nm and W: 100nm; (b) Calculated transmission spectrum with FDTD method for the plasmonic racetrack resonator; Inset: Schematic of the plasmonic racetrack resonator, brown region: metal, brown region: insulator. R: 500nm, L: 670nm, W: 100nm and G: 25nm.

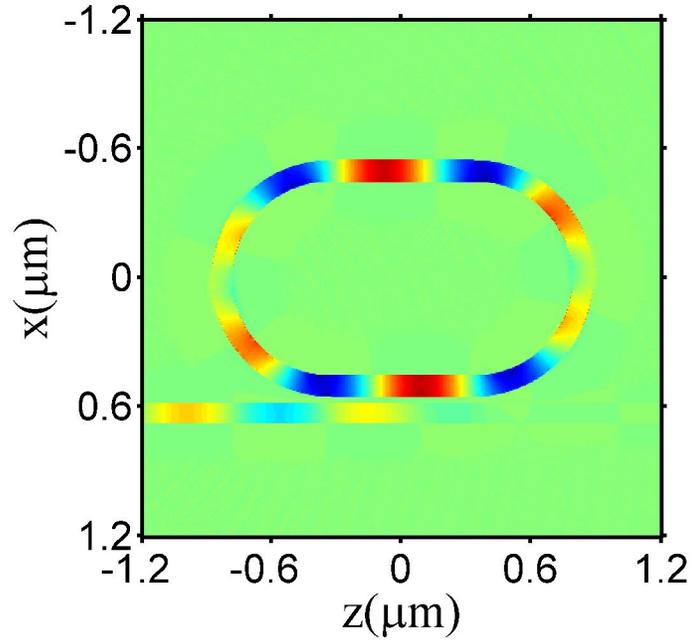

(a)

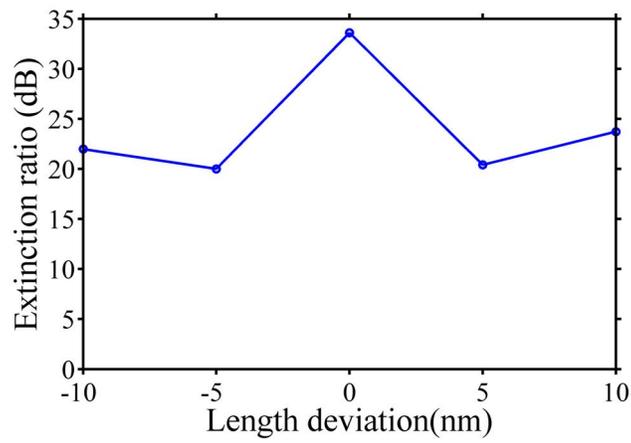

(b)

Fig. 4 (a) the magnetic field profile for the wavelength 1565nm; (b) Extinction ratios versus the length deviations of coupling region.